\documentclass[prd,reprint,superscriptaddress,preprintnumbers]{revtex4-1}
\pdfoutput=1
\usepackage{amsmath,amssymb,graphics}
\usepackage{verbatim}
\usepackage{graphicx}
\usepackage{color}
\usepackage{mathrsfs}
\usepackage{ragged2e}
\usepackage{float}

\usepackage{esint}
\usepackage[unicode=true,pdfusetitle,
bookmarks=true,bookmarksnumbered=false,bookmarksopen=false,
breaklinks=false,pdfborder={0 0 1},backref=false,colorlinks=true]
{hyperref}
\hypersetup{
	citecolor=black,linkcolor=black,urlcolor=black}

\def\equationautorefname~#1\null{eq.\,(#1)\null}

\usepackage[hang,flushmargin]{footmisc}
\allowdisplaybreaks
\makeatletter

\usepackage{etoolbox}
\apptocmd{\thebibliography}{\justifying\setlength{\leftskip}{7.4mm}}{}{}

\usepackage{relsize}

\makeatletter
\def\simgt{\mathrel{\lower2.5pt\vbox{\lineskip=0pt\baselineskip=0pt
			\hbox{$>$}\hbox{$\sim$}}}}
\def\simlt{\mathrel{\lower2.5pt\vbox{\lineskip=0pt\baselineskip=0pt
			\hbox{$<$}\hbox{$\sim$}}}}
\makeatother

\usepackage{changepage}

\newcommand{\be}{\begin{equation}}
	\newcommand{\ee}{\end{equation}}
\newcommand{\bea}{\begin{eqnarray}}
	\newcommand{\eea}{\end{eqnarray}}

\usepackage{color}
\definecolor{darkgreen}{rgb}{0,0.5,0}
\definecolor{darkred}{rgb}{0.5,0,0}
\definecolor{darkblue}{rgb}{0,0,0.6}
\definecolor{purple}{rgb}{0.4,.2,0.7}

\begin{document}
	
	\title{Cosmology with Higher-Derivative Gravities}
	
	\author{H. Khodabakhshi}
	\email{h\_khodabakhshi@tju.edu.cn}
	\affiliation{Center for Joint Quantum Studies and Department of Physics,\\
		School of Science, Tianjin University, Tianjin 300350, China \\}
	
	\author{M. Farhang}
	\email{m\_farhang@sbu.ac.ir}
	\affiliation{Department of Physics, Shahid Beheshti University, 1983969411, Tehran, Iran \\}
	
	\author{F. Shojai}
	\email{fshojai@ut.ac.ir}
	\affiliation{Department of Physics, University of Tehran, P.O. Box 14395-547, Tehran, Iran \\}
	
	\author{H. L\"u}
	\email{mrhonglu@gmail.com}
	\affiliation{Center for Joint Quantum Studies and Department of Physics,\\
		School of Science, Tianjin University, Tianjin 300350, China \\}
	\affiliation{{ Joint School of National University of Singapore and Tianjin University,\\ International Campus of Tianjin University, Binhai New City, Fuzhou 350207, China}}
	
	\begin{abstract}
We introduce an ingenious approach to explore cosmological implications of higher-derivative gravity theories. The key novelty lies in the characterization of the additional massive spin-0 modes constructed from Hubble derivatives as an effective density, with the corresponding pressure uniquely determined by energy conservation, while terms with no Hubble derivatives directly alter Friedmann equations. This classification of the various high-derivative contributions to Friedmann equations develops insight about their cosmological impacts and is essential for understanding the universe's evolution across energy scales. Various examples of higher-derivative gravity theories illustrate the power of this method in efficiently solving Friedmann equations and exploring new phenomena. Using CMB and BAO data, we apply this method to assess the observational feasibility of wall-bouncing universes, as predicted by scenarios with, e.g., certain third order modifications to general relativity. These models also provide an inflationary phase without the need to introduce extra scalar fields.
	\end{abstract}
	
	\maketitle
	\allowdisplaybreaks

\section{Introduction}

General relativity (GR) has passed numerous observational tests across a wide range of scales and plays a crucial role in understanding the large-scale universe. However, due to conceptual challenges and observational tensions associated with GR and $\Lambda$CDM, serious efforts have been made to explore its modifications. There are different ways of its modification, such as Modified Gravity (MOG) \cite{mog,fcb} or Massive Gravity \cite{massive}. Keeping the general covariance, higher-derivative gravity theories present one natural category within a broader range of modifications to GR. 
Einstein's gravity can be viewed as an effective low-energy theory that will receive higher-derivative corrections which become important as the energy scale increases, such as near black hole singularities and in the early universe \cite{Hooft}.
In this paper, we will focus on the cosmological implications of higher-derivative gravity theories \cite{h1,h2}. In this context, a general form of the Lagrangian is given by
$
\mathcal{L} = \sqrt{-g} L(g,\partial g, \partial^2 g, \ldots)
$.
Some higher-dimensional Lagrangians, such as Lovelock gravities \cite{Love}, can be expressed as the sum of polynomials of first-order derivative terms and total derivative terms as \cite{Pad1, Hossein1}
$
\mathcal{L} = \sqrt{-g} L_{\text{bulk}}(g,\partial g) + \partial_{\mu} J^\mu
$,
leading to second-order equations of motion. Nevertheless, higher-derivative gravity terms would generally lead to higher-order differential equations of motion and more complex dynamics with additional degrees of freedom. For instance, adding the quadratic curvature terms \(\alpha R^2 + \beta C^2\), where \(C^2\) is the square of the Weyl tensor, can describe a system with a massive spin-0 mode of \(m_\alpha^2 = 1/(2 \alpha)\) and a massive spin-2 mode of \(m_\beta^2 =1/(2\beta)\), distinct from the massless gravitons of GR \cite{Stelle:1976gc, Stelle:1977ry}. 

In the context of cosmology, when the ansatz is the FLRW metric, the scale factor \(a(t)\) and its derivatives describe global, isotropic changes in the scale of the universe, which are scalar in nature. Hence one may expect that the least contribution to the Friedmann equation from a scalar spin-0 mode modification to GR can be  characterized similar to a perfect fluid described by its effective density and  pressure. While in four dimensions the \(C^2\)-term will not affect the Friedmann equations since it does not directly lead to spin-0 modes \cite{m1,m2,m3, Hong1}. Furthermore, we expect the effective density coming from high-derivative corrections to be important at high-energy scales by providing an inflationary phase in the early universe. On the other hand, the high-derivative gravity terms also can leave \(a\) and \(\dot{a}\) terms in the effective form of the Lagrangian, which can affect the evolution of the universe at lower-energy scales (as we will show in $R^3$ and RGB models later). Moreover, for Lagrangians with the bulk term $L_{\text{bulk}}(g,\partial g)$, e.g., in Horndeski gravity \cite{Hor} such as four-dimensional Einstein-Gauss-Bonnet (4D-EGB) gravity \cite{Lin, Hong2, Hossein2}, or generalized quasi-topological gravity such as Einstein Cubic Gravity (ECG) \cite{Robert1, Pab1, Pab2, Hong3}, high-derivative gravity corrections effectively modify the first Friedmann equation as $H^2+\lambda H^4=\rho_{\text{tot}}(a)/3$ and $H^2-\lambda H^6=\rho_{\text{tot}}(a)/3$ in 4D-EGB and ECG respectively (note that in the framework of the 4D-Horndeski-type scalar-tensor theory, the scalar field in the 4D-EGB Lagrangian can play a radiation-like density role)\cite{jap,Hossein3, Hossein4}. In the context of high-derivative gravities, it is quite important to devise a method for dividing the role of each term in the Friedmann equations to provide an analysis of the expansion history and the inflationary phase of the universe.

Based on our discussions, in the context of cosmology with FLRW metric if the Lagrangian is high-derivative, i.e., $L_{\text{bulk}}(g,\partial g, \partial^2 g,...)$, then it may describe additional massive spin-0 modes, such as $R^2$ and $R^3$ modifications to GR, while a Lagrangian as $L_{\text{bulk}}(g,\partial g)$ will only alter the LHS of the first Friedmann equation, such as 4D-EGB and ECG. We also know the first Friedmann equation is actually a Hamiltonian constraint, and its derivative order is the same as the bulk Lagrangian. Therefore, we choose to keep the Hubble parameter (related to the first-order derivative of the metric in the bulk Lagrangian) on the LHS of the first Friedmann equation, while its derivatives are treated as an effective density on the RHS. Then, according to this new total density $\rho_{\text{tot}}=\rho+\rho_{\text{eff}}$, one can also calculate the unique effective pressure using the conservation of energy equation.

We examine the contribution of $\rho_{\text{eff}}$ across different cosmic eras to effectively solve the high-order differential Friedmann equations. We will show that the $\rho_{\text{eff}}$ contribution can be significant at high-energy scales, and it leads to constraints on the coupling constants. We use various examples to demonstrate the validity and power of this method in providing a clear picture of the universe's evolution and exploring new cosmological phenomena. For example, we will show that third-order modifications to GR, such as $R^3$ and RGB, lead to a wall-bounce universe even with a normal matter sector. In contrast with the standard bounce, a wall-bounce  scenario indicates that the universe originated from a bouncing point with a nonzero Hubble parameter at this point. Using CMB and BAO data, we modify the CosmoMC package\cite{cosmomc} to investigate the observational tests of the presented models.

Our paper is organized as follows. In Sec. \ref{2}, we explain the method of dividing different terms in the Friedmann equations based on their roles at different energy scales to define a consistent effective density and pressure. In Sec. \ref{3}, using various examples and employing our new method, we investigate the expansion history of the hot phase of the universe. In Sec. \ref{4}, using CMB and BAO data, we investigate the observational feasibility of the presented examples. In Sec. \ref{5}, we study the contribution of the effective density and pressure at high-energy scales and the possibility of the inflationary phase in different examples without adding an extra scalar field. In the last section, we conclude our paper, and in the Appendix, we explain that the RGB model is actually the first-order correction of the Gauss-Bonnet extended Starobinsky gravity.

\section{Method}\label{2}

The effective Lagrangian leads to the Friedmann equations by taking the variation with respect to the lapse function \(N\) and \(a\), where we have
$
ds^2 = -N(t)^2 dt^2 + a(t)^2 (dx_1^2 + dx_2^2 + dx_3^2).
$
 Setting \(N = 1\) gives the first Friedmann equation, which is equivalent to the Hamiltonian constraint, and the acceleration equation, respectively. These can be expressed in a general form in terms of \(H = \dot{a}/a\) as
\begin{align} \label{f1}
	&f(H) + g(\dot{H}, H \dot{H}, \ldots) = \frac{\rho}{3},
	\\ \label{f2}
	&h(H) + k(\dot{H}, H \dot{H}, \ldots) = -p,
\end{align}
where \(\rho = \rho_{\rm m} + \rho_{\rm r} + \Lambda\), \(p = p_{\rm m} + p_{\rm r} - \Lambda\) describe a perfect fluid with the energy-momentum tensor \(T^\mu_\nu = \text{diag}(\rho, p, p, p)\) that satisfies the conservation of energy equation \(\nabla_{\mu} T^{\mu}_{\nu} = 0\), and is minimally coupled to the Lagrangian as \(\mathcal{L} - V(a)\) through the potential term
\begin{equation}
	\rho = \frac{V(a)}{2a^3}, \quad p = -\frac{V'(a)}{6a^2}.
\end{equation}
The order of the derivatives in Eq. (\ref{f1}) is the same as in the bulk term of the Lagrangian. This gives us a clue to separate different roles of higher derivative terms appropriately. The \(\dot{H}\) (or \(\ddot{a}\)) and its higher order derivatives appearing in Eq. (\ref{f1}) can be considered as an effective density associated with the scalar modes of the universe geometry. Thus, the equations (\ref{f1}) and (\ref{f2}) can be written as
\begin{align} \label{fe1}
	&f(H) = \frac{\rho_{\text{tot}}}{3},
	\\
	&u(H) + z(\dot{H}, H \dot{H}, \ldots) = -p_{\text{tot}}.
\end{align}
where \(\rho_{\text{tot}} = \rho + \rho_{\text{eff}}\), \(\rho_{\text{eff}} = -g(\dot{H}, H \dot{H}, \ldots)\), \(p_{\text{tot}} = p + p_{\text{eff}}\), and the effective pressure \(p_{\text{eff}}\) is uniquely obtained using the conservation of energy equation
$
\dot{\rho}_{\text{tot}} + 3H(\rho_{\text{tot}} + p_{\text{tot}}) = 0.
$
The  equation of state parameter is also defined as 
$
\omega_{\text{tot}}= p_{\text{tot}}/\rho_{\text{tot}}.
$
Here the method to determine  \(\rho_{\text{eff}}\) is quite unique to guarantee consistency with observational tests across different cosmic eras.
 We consider the terms \(\dot{H}\) and its higher order derivatives as the effective density and expect these terms to be important at the high energy scales such as the inflationary era, where we consider slow-roll conditions \cite{sllowrol} by keeping terms up to the first-order in \(\dot{H}\). In our scenario, the evolution of the universe is typically represented by the standard density \(\rho\). We expect the contribution of \(\rho_{\text{eff}}\) to be negligible during the hot phase of the universe, and Eq. (\ref{fe1}) gives the expansion of the universe as
\begin{equation}
H(z) = f^{-1}(\rho(z)/3).
\end{equation}
Substituting \(H(z)\) from the above equation into \(\rho_{\text{eff}}\), considering \(\rho(z) = C_n (1+z)^n\) (\(n = 0, 3, 4\) correspond to cosmological constant, matter and radiation dominated eras, respectively), we will examine through various examples whether \(\rho_{\text{eff}}\) is actually negligible as we supposed or not. This contribution depends on the order of metric derivatives  appearing in the bulk Lagrangian of the high-derivative correction terms, and it leads to constraints on the coupling constants. For example, as we will  see, the effective density  associated to \(R^2\) and \(R^3\)  is proportional to \((1+z)^{2n}\) and \((1+z)^{3n}\), respectively. 
The requirement that 
$\rho_{\text{eff}} \ll \rho$ in the hot (i.e., non-inflationary) phase of the Universe imposes constraints on the coupling constants in the corresponding models. 
On the other hand, \(\rho_{\text{eff}}\) is constructed from higher-order derivatives, expected to play role at high energies which hints to the possibility of its dominant role in a pre-radiation era. It is therefore intriguing to explore the possible inflationary behaviour of \(\rho_{\text{eff}}\) at early times. 
	
\section{Hot phase of the Universe}\label{3}

Using the approach we provided for defining effective density and pressure to understand the role of high-order derivative terms, we will consider various examples to investigate the expansion history of the hot phase of the Universe. In this phase we first assume $\rho_{\text{eff}}$ is negligible compared to \(\rho(z)= C_n (1+z)^n\) and only 
consider the impact of the modifications to the first Friedmann equation in terms of non-high derivative terms (LHS of Eq.~(\ref{fe1})).
We then investigate the consistency of this assumption. 

\subsection{$R^2$, $C^2$, and $C^3$}

Consider the Lagrangian
\begin{equation}
\mathcal{L}_{R^2} = \sqrt{-g} (R + \alpha R^2),
\end{equation}
known as Starobinsky gravity within the framework of \(f(R)\)-gravity \cite{Starobinsky, fr, fr1,  fr2, fr3}. The Friedmann equations are
\begin{align}\label{er2}
	&H^2 = \frac{\rho_{\text{tot}}}{3},\\ \label{er21}
	&3H^2 + 2\dot{H} = -p_{\text{tot}},
\end{align}
with effective density and pressure contributions as
\begin{align}\label{dr2}
	\rho_{\text{eff-}R^2} &= 18 \alpha \left(-6 H^2 \dot{H} + \dot{H}^2 - 2 H \ddot{H} \right), \\
	p_{\text{eff-}R^2} &= 6 \alpha \left(18 H^2 \dot{H} + 9 \dot{H}^2 + 12 H \ddot{H} + 2 \dddot{H} \right).
\end{align}
From Eq.~(\ref{er2}) and assuming $\rho_{\text{eff}} \approx 0$ we find \(H(z)^2 \approx C_n (1+z)^n/3\). Substituting \(H(z)\) into the effective density term yields
\begin{equation}
\rho_{\text{eff-}R^2}(z) = \frac{3}{2} \alpha C_n^2 n (4-n) (1+z)^{2n}.
\end{equation}
This shows that effective density can satisfy the weak energy condition (WEC) for \(n < 4\), with zero contributions for \(n = 0\) and \(n = 4\). 
In the matter-dominated era, assuming \(\alpha \lessapprox 10^{-9} M_{\rm Pl}^{-2}\) \cite{Planck2018} would  guarantee that  
$\rho_{\text{eff}} \ll \rho$, or equivalently, $(1+z)^{3} \ll 1/(\alpha C_3)$, across the whole redshift range of interest. Therefore, the post-inflationary expansion of the Universe is given by the standard cosmological model \cite{P18}, where $\rho_{\text{eff}}$ and $p_{\text{eff}}$ are negligible in Eqs. (\ref{er2}) and (\ref{er21}), while  they are expected to play a more significant role in the  high-energy inflationary Universe.

Two other notable cases are
\begin{align}
&\mathcal{L}_{C^2} = \sqrt{-g} (R + \alpha C^2),\\
&\mathcal{L}_{C^3} = \sqrt{-g} (R + \alpha R^2 + \beta C^3),
\end{align}
with \(C^2 = C_{\alpha \beta \gamma \sigma} C^{\alpha \beta \gamma \sigma}\) and two forms of \(C^3\) as:
$
C_{(1)}^3 = C^{\alpha \beta \gamma \sigma} C_{\alpha \eta \iota \beta} C^{\eta \ \ \iota}_{\ \gamma \sigma},
$
$
C_{(2)}^3 = C^{\alpha \ \gamma}_{\ \beta \ \sigma} C^{\beta \ \iota}_{\ \gamma \ \eta} C^{\sigma \ \ \eta}_{\ \alpha \iota},
$
where $C_{\alpha \beta \gamma \sigma}$ is the Weyl tensor. Since these terms describe a massive spin-2 mode \cite{Hong1,Hooft}, they do not affect the Friedmann equations in four dimensions. They might, however, influence the cosmological model when considering perturbed equations.

\subsection{\(R \Box R\)}

Another interesting example involves a Lagrangian incorporating derivatives of the Ricci or Riemann tensor. Consider
\begin{equation}
	\mathcal{L}_{R\Box R} = \sqrt{-g} (R + \alpha R^2 + \beta R \Box R).
\end{equation}
This model leads to Friedmann equations similar to those in Eqs. (\ref{er2}) and (\ref{er21}), where the effective density and pressure are
\begin{widetext}
	\begin{align} \label{dRBR}
		\rho_{\text{eff-}R\Box R} =  - 18 \beta \bigg( 24 H^4 \dot{H} + 8 \dot{H}^3 - 10 H^3 \ddot{H}  + 2 \dot{H} H^{(3)}  - \ddot{H}^2 - 4 H^2 \left( 8 \dot{H}^2 + 3 H^{(3)} \right) - 2 H \left( 12 \dot{H} \ddot{H} + H^{(4)} \right) \bigg), \nonumber \\
		p_{\text{eff-}R\Box R} = + 6 \beta \bigg( 72 H^4 \dot{H}  - 6 H^3 \ddot{H} - 27 \ddot{H}^2 - 46 H^2 H^{(3)} - 40 \dot{H}^3 - 42 \dot{H} H^{(3)} - 2 H \left( 83 \dot{H} \ddot{H} + 9 H^{(4)} \right) - 2 H^{(5)} \bigg),
	\end{align}
\end{widetext}
and we have \(\rho_{\text{eff}} = \rho_{\text{eff-}R^2} + \rho_{\text{eff-}R\Box R}\), \(p_{\text{eff}} = p_{\text{eff-}R^2} + p_{\text{eff-}R\Box R}\). By considering \(\rho (z) = C_n (1+z)^n\), we find the effective density as
\begin{equation}
	\rho_{\text{eff-}R\Box R}(z) = \frac{1}{3} \beta C_n^3 (4 - n) n (-6 - 8n + 5n^2) (1 + z)^{3n}.
\end{equation}
For \(2 < n < 4\), the effective density satisfies WEC, and it shows that the contribution from \(R \Box R\) to the density \(\rho\) is zero for \(n=0\) and \(n=4\). In the case of \(n=3\), \(\rho_{\text{eff-}R\Box R}\) is negligible if \(\beta\) satisfies the constraint \(\beta \ll (1+z)^{-6}/C_3^2\) in the matter dominated era, which is feasible as \(\beta\) is considered a third-order correction, expected to satisfy \(\beta \ll M_{\rm Pl}^{-4}\). Therefore, \(\rho_{\text{eff-}R\Box R}\)'s impact is insignificant during the hot phase of the Universe but may contribute during the inflationary phase. After inflation, the expansion history is given by the standard cosmological model \cite{P18} similar to the $R^2$ model.
	
\subsection{\(R^3\) and RGB}

In this example, we consider third-order corrections to Einstein gravity via the Lagrangians:
\begin{equation}
\mathcal{L}_{R^3} = \sqrt{-g} (R + \alpha R^2 + \beta R^3)
\end{equation}
and
\begin{equation}\label{LRGB}
\mathcal{L}_{RGB} = \sqrt{-g} (R + \alpha R^2 + \beta \text{RGB}),
\end{equation}
where \(\text{RGB} = R \times (R^2 - 4 R_{\mu \nu} R^{\mu \nu} + R_{\mu \nu \rho \sigma} R^{\mu \nu \rho \sigma})\).  The RGB model is  the first-order extended GB-correction to the Starobinsky Lagrangian (see appendix \ref{app1}). These models yield Friedmann equations:
\begin{align}\label{er3}
	&H^2 - \lambda_i H^6 = \frac{\rho_{\text{tot}}}{3},\\\label{er31}
	&3H^2 + 2\dot{H} - 3\lambda_i (H^6 + 2H^4\dot{H}) = -p_{\text{tot}},
\end{align}
with \(\lambda_1 = 144 \beta\) and \(\lambda_2 = 24 \beta\) for \(R^3\) and RGB, respectively. The modified Friedmann equations (\ref{er3}) and (\ref{er31}) satisfy the energy conservation equation and result in
\begin{widetext}\label{dR3}
	\begin{align}
		&\rho_{\text{eff-}R^3} = - \frac{3\lambda_1}{4} \left( \dot{H} \left( 36 H^4 + 15 H^2 \dot{H} - 2 \dot{H}^2 \right) + 6 H \left( 2 H^2 + \dot{H} \right) \ddot{H} \right),
		\nonumber \\
		&p_{\text{eff-}R^3} = + \frac{3 \lambda_1}{4} \left( 36 H^4 \dot{H} + 63 H^2 \dot{H}^2  + 24 H^3 \ddot{H} + 8 \dot{H}^3+ 28 H \dot{H} \ddot{H} + 2 \ddot{H}^2 + 2 \left(2 H^2 + \dot{H}\right) H^{(3)} \right),\nonumber \\
   		&\rho_{\text{eff-RGB}} = - 3 \lambda_2 H^2 \left(6 H^2 \dot{H} + \dot{H}^2 + 2 H \ddot{H} \right),	\nonumber \\
		&p_{\text{eff-RGB}} =  + \lambda_2 \left(18 H^4 \dot{H}  + 12 H^3 \ddot{H} + 8 H \dot{H} \ddot{H}+ 2 \dot{H}^3 + H^2 \left(27 \dot{H}^2 + 2 H^{(3)} \right)\right). \label{dRGB}
	\end{align}
\end{widetext}
in the case of RGB. Eqs. (\ref{er3}) and (\ref{er31}) show that the $R^3$ and RGB models can be classified in the same category as they have the same form of their Friedmann equations, which are also similar to the ECG model where $\rho_{\text{tot}} = \rho$ and $p_{\text{tot}} = p$ \cite{Hong3}. The solutions to the Friedmann equation (\ref{er3}) are 
\begin{align}\label{sol1}
	&H^2 = \frac{2}{\sqrt{\lambda_i}} \cos\left(\frac{1}{3} \left[ \arccos\left(\frac{\rho_{\text{tot}} \sqrt{\lambda_i}}{2}\right) + \pi\right] \right)
	\lambda_i > 0,
	 \\
	&H^2 = \frac{2}{\sqrt{|\lambda_i}|} \sinh\left(\frac{1}{3} \arcsin\left(\frac{\rho_{\text{tot}} \sqrt{|\lambda_i|}}{2}\right) \right) \quad \lambda_i < 0.\label{sol2}
\end{align}
The vacuum solution $\rho = 0$ leads to: $H = 0$, and $H^2 = 1/\sqrt{\lambda_i}$. For $\lambda_i > 0$ at $z = z_0$, where $H^2 = 1/\sqrt{\lambda_i}$, the density is
\begin{align}\label{bp}
	\rho_{\text{tot}}(z_0) = 2/\sqrt{\lambda_i}.
\end{align}
Near \(z_0\), the solution (\ref{sol1}) indicates that for matter satisfying the null energy condition (NEC) (\(w_{\text{tot}} \ge -1\)), we require \(z \le z_0\), suggesting a wall-bounce at \(z = z_0\). The mechanism at the bounce point \(z_0\) resembles a ping pong ball hitting a brick wall, contrasting with the traditional concept of a bounce where \(H=0\) at the bounce point \cite{Hong3, Hossein3, Hossein4}. The Hubble parameter for both \(\lambda_i>0\) and \(\lambda_i<0\) would converge to
\begin{equation}
	H^2 = \rho_{\text{tot}}/3 + \lambda_i \left(\rho_{\text{tot}}/3\right)^3/3 + ...\, ,
\end{equation}
as the dimensionless parameter \(Q = \sqrt{\lambda_i} \rho_{\text{tot}}/2\) approaches zero and it implies the Einstein limit for \(\rho_{\text{eff}} = 0\).
	
	Similar to the previous example, \(\rho_{\text{eff}}\) for these models, assuming \(\rho(z) = C_n (1+z)^n\) and keeping terms up to the first-order of \(\lambda_i\) for simplicity, can be written as
	\begin{equation}
	\rho_{\text{eff-}R^3} (z) \approx \frac{1}{144} \lambda_1 C_n^3 \, n (n-3) (-24 + 5n) (1 + z)^{3n}  +...\, ,
	\end{equation}
	 and
	\begin{equation}
	\rho_{\text{eff-RGB}}(z) \approx
	\frac{1}{36} \lambda_2 C_n^3 n \left(12 - 5n \right) (1 + z)^{3n} +...\,.
	\end{equation}
These equations show $\rho_{\text{eff-}R^3}$ is zero for $n=0$, $n=3$ and it violates WEC for $n=4$
	, and also
	$\rho_{\text{eff-RGB}}(z)$, is zero for $n=0$, and it violates WEC for $n=3$, $n=4$. Following same logic as the $R^2$ and $R \Box R$ models, these calculations confirm that \(\rho_{\text{eff}}\) is negligible during lower-energy epochs, but it can be significant during the inflationary phase.
	
\section{Observational Feasibility}\label{4}

In the case of $R^2$ and $R \Box R$ models, the expansion history during the hot universe is given by the Friedmann equations (\ref{er3}) and (\ref{er31}), where the effective density and pressure are negligible. The cosmological parameters are then expected to agree with the measurements of the standard, GR-based cosmological scenario \cite{P18}. Regarding the case of $R^3$ and RGB models, the solutions are provided by Eqs. (\ref{sol1}) and (\ref{sol2}). Given the observational success of Einstein gravity, one may require that deviations from Einstein gravity in the $R^3$ and RGB models be small. This can be ensured by choosing sufficiently small values for $\lambda_i$. Both wall-bouncing ($\lambda_i > 0$) and non wall-bouncing ($\lambda_i < 0$) scenarios are highly sensitive to the choice of $\lambda_i$ (non wall-bounce indicates no bounce). The reason is that in these models, we observe \( H^2/(1+z)^4 \rightarrow 0\), whereas for the Einstein case, \( H^2/(1+z)^4 \rightarrow H_0^2 \Omega_\text{r} \) as \( z \rightarrow \infty \). Thus, the value of \(\lambda_i\) should be small enough to respect the thermal history of the Universe. In the wall-bouncing scenario, the value of \(\lambda_i\) can determine the redshift of the wall-bouncing point, as shown in Eq.~(\ref{bp}). For instance, if we require a wall-bounce to happen before the lepton era, with \(z \sim 10^{15}\), and assuming that the \(\Lambda\)CDM parameters are set to their $Planck$ bestfits \cite{P18}, we find \(\lambda_i \lessapprox 10^{-120} (\text{km/s/Mpc})^{-2}\).
To compare the predictions of the  \(R^3\) and RGB models for the expansion history with observations and constrain their parameters, we integrate these models into the CosmoMC package \cite{cosmomc}, originally developed for  \(\Lambda\)CDM, as we have done in \cite{Hossein4} for the 4D-EGB gravity. We also  assume that the bounce, if any, occurred before \(z_{\text{bounce}} \gtrsim 10^{10}\). As the datasets we use the  anisotropy spectrum of the cosmic microwave background radiation and its lensing from $Planck$ 2018  (TT+TE+EE+lensing), along with baryonic acoustic oscillations (BAO) measurements, referred to as P18+BAO. Table~\ref{T1} summarizes the results. In brief, the results are found to be consistent with GR. In the next section,we will investigate the possibility of an early inflationary scenario in these model.
	
\begin{table}[ht]
		\centering
		\caption{Constraints on the parameters of the \(R^3\) and RGB models for non wall-bouncing  (\(\lambda_i<0\)) and wall-bouncing universes (\(\lambda_i>0\)) measured by P18+BAO.}
		\begin{tabular}{@{}cccc@{}}
			\hline \hline
			\multicolumn{2}{c}{$\hspace{1.7cm}$\(\lambda_i>0\)} & \multicolumn{2}{c}{\(\lambda_i<0\)} \\
			\hline
			\(\tiny \Omega_{\rm b} h^2\) & \(\tiny 0.02241\pm0.00026\) & \(\tiny 0.02240\pm0.00027\) & \\
			\(\tiny \Omega_{\rm c} h^2\) & \(\tiny 0.1195\pm0.0017\) & \(\tiny 0.1195\pm0.0018\) & \\
			\(\tiny H_0\) & \(\tiny 67.56\pm0.82\) & \(\tiny 67.58\pm0.83\) & \\
			\(\tiny |\lambda_i|\) & \(\tiny 10^{-59}\) & \(\tiny 10^{-34}\) & \\
			\(\tiny \tau\) & \(\tiny 0.057\pm0.015\) & \(\tiny 0.057\pm0.014\) & \\
			\(\tiny \ln(10^{10}A_s)\) & \(\tiny 3.049\pm0.028\) & \(\tiny 3.049\pm0.030\) & \\
			\(\tiny n_{\rm s}\) & \(\tiny 0.9660\pm0.0069\) & \(\tiny 0.9659\pm0.0070\) & \\
			\hline
			\(\text{log}(|\lambda_i| H_0^2)\) & \(\tiny <-55\) & \(\tiny <-30\) & \\
			\(\tiny \sigma_8\) & \(\tiny 0.811\pm0.012\) & \(\tiny 0.811\pm0.012\) & \\
			\hline
		\end{tabular}
		\label{T1}
	\end{table}
	
\section{Inflationary phase}\label{5}

We found that \(\rho_{\text{eff}}\) contributes minimally during the hot phase of the Universe. However, higher derivative terms in the Lagrangian can be crucial for driving the accelerated expansion during the inflationary era.  Now we will investigate the  behavior of the presented examples in the context of slow-roll inflation, where $\epsilon \equiv -\dot{H}/H^2 \ll 1$ is the slow-roll parameter.
	
\subsection{\(R^2\), \(C^2\), and \(C^3\)}

Imposing slow-roll conditions on the effective density (\ref{dr2}), the Friedmann equation (\ref{er2}) simplifies to
\begin{equation}
H^2 \approx \rho_{\text{inf}}/3, \quad \rho_{\text{inf}}/3 \approx - 36 \alpha H^2 \dot{H},
\end{equation}
where \(\alpha = 1/(6 m_{\alpha}^2)\) is linked to the scalaron mass or the massive scalar mode \cite{Planck2018,value}. The condition \(\ddot{a} > 0\) ensures inflation where in this case we have \(\omega_{\text{eff}} \approx -1\). Inflation ends when \(\epsilon(t_\text{f}) \approx 1\). The number of e-folds \(\mathcal{N}\) is:
	$
	\mathcal{N} = \int_{H_\text{i}}^{H_\text{f}} \left(H/\dot{H}\right) dH.
	$
	Defining the boundaries of integration, with \(t = t_\text{i}\) at the onset and \(t = t_\text{f}\) at the end of inflation, we have
	$
	H_\text{i} = \sqrt{-\dot{H}/\epsilon_\text{i}} \big|_{t = t_\text{i}}, H_\text{f} = \sqrt{-\dot{H}} \big|_{t = t_\text{f}}
	$. For addressing the flatness and horizon issues, we require \(\mathcal{N} \approx 60\), which gives \(\epsilon_\text{i} = 8.2 \times 10^{-3}\).
	
	\subsection{\(R\Box R\)}
	
	In this case, applying slow-roll conditions on the effective density (\ref{dRBR}), the Friedmann equation (\ref{er2}) can be expressed as
	\begin{equation}
		H^2 \approx \rho_{\text{inf}}/3, \quad \rho_{\text{inf}}/3 \approx - 36 \alpha H^2 \dot{H} -144 \beta H^4 \dot{H},
	\end{equation}
	where we have two massive scalar modes with mass-squared \(m_\alpha^2 = 1/(2\alpha)\) and \(m_\beta^2 = 1/(2\beta)\). The condition for accelerated expansion is  again met with \(\omega_{\text{eff}} \approx -1\). Also \(\mathcal{N}\) is calculated as
	\begin{equation}
		\mathcal{N} = -18 H^2 (\alpha + 2 \beta H^2) \big|_{H_\text{i}}^{H_\text{f}} ,
	\end{equation}
	in the same approach we used for the \(R^2\) model, where we have \(H_i = H|_{\epsilon = \epsilon_i}\) and \(H_f = H|_{\epsilon \approx 1}\) in the following equation
	\begin{align}\label{hrbr}
		H = \left(18 \alpha \epsilon + 6 \sqrt{\epsilon (4 \beta + 9 \alpha^2 \epsilon)}\right)^{-\frac{1}{2}}.
	\end{align}
	Up to the first-order of \(\beta\), \(\mathcal{N}\) can be obtained as
	\begin{equation}
		\mathcal{N} = \frac{1}{2} \left(-1 + \frac{1}{\epsilon_\text{i}}\right) -\frac{1}{36 \alpha^2} \left(-1 + \frac{1}{\epsilon_\text{i}^2}\right)\beta + \ldots.
	\end{equation}
	Similar to the \(R^2\) case, given that \(\mathcal{N} \approx 60\), \(\alpha \approx 10^{-9} M_{\rm Pl}^{-2}\) \cite{Planck2018, value}, and choosing \(\epsilon_{\text{i}} \approx 4.1 \times 10^{-3}\) results in \(\beta \approx 10^{-11} M_{\rm Pl}^{-4}\). Note that decreasing $\beta$ to, e.g., \(10^{-22} M_{\rm Pl}^{-4}\) would increase $\epsilon_{\text{i}}$ to approximately \( 8.2 \times 10^{-3}\).
	
	\subsection{\(R^3\) and RGB} 
	
	Imposing slow-roll conditions on the effective densities in Eq. (\ref{dRGB}), the Friedmann equation (\ref{er3}) yields
    \begin{equation}
	H^2 - \lambda_i H^6 \approx \rho_{\text{inf-i}}/3,
	\end{equation}
	where we have
	\begin{align}
	&\rho_{\text{inf-1}}/3 \equiv \rho_{\text{inf-}R^3}/3 \approx -9 (4 \alpha + \lambda_1 H^2) H^2 \dot{H},\\
	&\rho_{\text{inf-2}}/3 \equiv \rho_{\text{inf-RGB}}/3 \approx -6 (6 \alpha + \lambda_2 H^2) H^2 \dot{H}.
	\end{align}
	These models give two massive scalar modes with mass-squared \(m_\alpha^2 = 1/(2\alpha)\) and \(m_\beta^2 = 1/(2\beta)\), where \(\beta\) can be determined by \(\lambda_i\). For a non wall-bouncing universe with \(\lambda_i < 0\) (or \(\beta < 0\)), this implies a negative mass-squared associated with a tachyon mode, potentially leading to tachyonic instability. To assess the accelerated expansion during inflation, we calculate the acceleration, \(\ddot{a}\), using:
	\begin{align}
		\frac{\ddot{a}}{a} =& \frac{4 \lambda_i H^6 - \rho_{\text{inf-}i} \left(\frac{1}{3} + \omega_{\text{eff-}i}\right)}{2 \left(1 - 3 \lambda_i H^4\right)}\nonumber \\& = \frac{2 \lambda_i H^6 + 9 \epsilon H^4 \left(4 \alpha + \lambda_i H^2\right)}{1 - 3 \lambda_i H^4}.
	\end{align}
	For \(\lambda_i < 0\) the condition
	\begin{align}
	H^2 < 36 \alpha \epsilon/((2 + 9 \epsilon) |\lambda_i|),
	\end{align}
	ensures \(\ddot{a} > 0\). For \(\lambda_i > 0\), the Universe starts from a wall-bounce at \(H^2 = 1/\sqrt{\lambda_i}\) and requires
	\begin{equation}
	H^2 < 1/\sqrt{3 \lambda_i},
	\end{equation}
	indicating post-bounce inflation. Now let's calculate \(\mathcal{N}\), for \(R^3\) and RGB models, using the same approach presented for the $R^2$ case.  For the $R^3$ model we have
	\begin{align}
	\mathcal{N} = \frac{3}{2} \left( \log(1 - \lambda_1 H^4) - \frac{8 \alpha}{\sqrt{\lambda_1}} \text{arctanh} \left(H^2 \sqrt{\lambda_1}\right) \right)\bigg|_{H_{\rm i}}^{H_{\rm f}},
	\end{align}
where  \(H_\text{i} = H|_{\epsilon = \epsilon_\text{i}}\) and \(H_\text{f} = H|_{\epsilon \approx 1}\) can be obtained form the following equation
\begin{align}\label{1b1}
	H = \left(18 \alpha \epsilon + \sqrt{\lambda_1 + 9 \epsilon (36 \alpha^2 \epsilon + \lambda_1)}\right)^{-\frac{1}{2}}.
\end{align}
Keeping the terms up to the first-order of \(\lambda_1\) gives
\begin{align}\label{Nf1}
	\mathcal{N} \approx \frac{1}{2} \left(-1 + \frac{1}{\epsilon_\text{i}}\right) - \frac{(4+ 27 \epsilon_\text{i} - 31 \epsilon_\text{i}^3)}{15552 \alpha^2 \epsilon_{\rm i}^3} \lambda_1 + \ldots,
\end{align}
indicating the \(R^3\) correction to the Starobinsky gravity. Regarding the RGB model, one can repeat the calculations that lead to
\begin{align}\label{N2}
	\mathcal{N} = \frac{3}{2} \left( \log(1 - \lambda_2 H^4) - \frac{12 \alpha}{\sqrt{\lambda_1}} \text{ArcTanh} \left[H^2 \sqrt{\lambda_2}\right] \right)\bigg|_{H_\text{i}}^{H_\text{f}},
\end{align}
where we have 
\begin{align}\label{hRGB}
	H = \left(18 \alpha \epsilon + \sqrt{\lambda_2 + 6 \epsilon (54 \alpha^2 \epsilon + \lambda_2)}\right)^{-\frac{1}{2}},
\end{align}
with \(H_\text{i} = H|_{\epsilon = \epsilon_\text{i}}\), \(H_\text{f} = H|_{\epsilon \approx 1}\) and expansion up to the first-order of \(\lambda_2\) yields
\begin{align}\label{Nf2}
	\mathcal{N} \approx \frac{1}{2} \left(-1 + \frac{1}{\epsilon_\text{i}}\right) - \frac{(2+ 9 \epsilon_\text{i} + 10 \epsilon_\text{i}^3)}{7776 \alpha^2 \epsilon_\text{i}^3} \lambda_2 + \ldots.
\end{align}
	It should be noted that during inflation, \(H_\text{i}\) and \(H_\text{f}\) must satisfy the positive acceleration conditions for both scenarios. In the case \(\lambda_i > 0\), considering \(\mathcal{N} \approx 60\) and substituting \(\lambda_i \approx 10^{-80} (\text{km/s/Mpc})^{-2}\) (which implies choosing a higher redshift for the wall-bouncing point from Eq. (\ref{bp})) and \(\alpha \approx 10^{-9} M_{\rm Pl}^{-2}\) \cite{Planck2018, value}, results in \(\epsilon_{\text{i}} = 8.4 \times 10^{-3}\). Choosing smaller values for \(\lambda_i\), such as \(10^{-120}\), does not affect the result, and the condition for positive acceleration is also completely satisfied. For the case of \(\lambda_i < 0\), even considering \(|\lambda_i| \approx 10^{-60} (\text{km/s/Mpc})^{-2}\), which is significantly smaller than the value of $\lambda_i$ from Table \ref{T1}, leads to an imaginary \(\mathcal{N}\). This result arises from the instability of the tachyon mode.
	
	\section{Conclusion}
	
	In this paper, within the framework of higher-derivative gravity theories and assuming the FLRW  as the spacetime metric, we proposed a novel approach to describe the role of the derivatives of the scale factor in Friedmann equations based on the way they impact the evolution of the universe. We showed that high derivative terms would only modify the dynamics of the universe in gravitational theories with scalar modes, while spin modes (in, e.g.,  \(C^2\) and \(C^3\) models in four dimensions) had no influence on Friedmann equations. We effectively associated to certain scalar modes, constructed from higher-derivative terms, an effective description in terms of a scalar field, characterized by  density and pressure similar to a perfect fluid. This scalar field was illustrated to have the potential to lead to an early inflationary phase in the universe. The other terms (with no Hubble derivatives) would alter the Friedmann equations and potentially impact the expansion history of the hot, non-inflationary phase of the cosmos. 
	
	Our approach provides a clear and effective method toward solving the nonlinear higher-order Friedmann equations. Various examples were employed to illustrate the validity and the power of this method. We showed that some models, such as $R \Box R$, contributed only a correction to $R^2$ during the inflationary phase, while $R^3$ and RGB could change the expansion history as well. These models provide two different scenarios: a wall-bounce and a non wall-bounce universe, depending on the sign of the coupling constant, where the negative sign would lead to a tachyonic mode. However, the theoretical appeal of these theories, especially in the wall-bouncing scenario where the initial singularity can be avoided, is encouraging for further investigation.
	
    We also explored the observational implications of these models. 
    In the case of $R^2$ and $R \Box R$ models, the form of the Friedmann equations is the same as in GR during the hot phase of the universe, and we expect the cosmological parameters to roughly agree with the measurements of the standard, GR-based cosmological scenario, where the coupling constants are constrained by inflation. For the case of $R^3$ and RGB models, we contrasted the predictions of these models against CMB and BAO observations and put upper bounds on the model parameters (see Table \ref{T1}) for both wall-bouncing and non-wall-bouncing scenarios. The bound on $\lambda_i$ in the bouncing case is mainly determined by the requirement of an early-enough bounce, if it exists, occurring before the highest redshift used in our analysis ($z_{\text{bounce}} \gtrsim 10^{10}$). This is a conservative choice, as a higher bounce redshift (and thus a lower $\lambda_i$) is required to respect the thermal history of the universe with an energy scale corresponding to the standard model of particle physics.
	
    Our study extends the potential for new investigations, providing an effective way for future theoretical and observational researches to explore novel cosmological aspects of high-derivative gravity theories.
	
\section*{Acknowledgement}

This work was supported in part by NSFC (National Natural Science Foundation of China) Grants No.~11935009 and No.~12375052.
	
\appendix
\section{RGB \& Gauss-Bonnet Extended Starobinsky Gravity}\label{app1}

Starobinsky gravity is the simplest example of $f(R)$ gravities with \( f(R) = R + \alpha R^2 \) and functions as an effective scalar-tensor theory where the scalar equation is algebraic. One can extend the scalar-tensor form of Starobinsky gravity with the Gauss-Bonnet invariant as
\begin{equation} \label{lag}
	\mathcal{L} = \sqrt{-g} \left(R + \phi R - \frac{1}{2} \mu^2 \phi^2 + U(\phi) \mathcal{G}\right),
\end{equation}
where
\begin{equation}
	\mathcal{G} = R^2 - 4 R_{\mu \nu} R^{\mu \nu} + R_{\mu \nu \rho \sigma} R^{\mu \nu \rho \sigma}, \hspace{0.5cm} \mu^2 = \frac{1}{2\alpha}.
\end{equation}
Taking the variation of the Lagrangian (\ref{lag}) with respect to \( g^{\mu \nu} \) yields the extended Einstein equation
\begin{widetext}\label{RGB}
\begin{align}
	&R_{\mu\nu} - \frac{1}{2}Rg_{\mu\nu} + \phi R_{\mu\nu} - \nabla_{\mu}\nabla_{\nu}\phi + g_{\mu\nu}\phi - \frac{1}{2}\phi Rg_{\mu\nu} + \frac{1}{4}\mu^2\phi^2 g_{\mu\nu} - 2R\nabla_{\mu}\nabla_{\nu}U \nonumber \\
	&-4\left(R_{\mu\nu}-\frac{1}{2}Rg_{\mu\nu}\right)\phi U + 8R_{\rho(\mu}\nabla_{\nu)}\nabla^{\rho}U - 4R_{\rho\sigma}\nabla^{\rho}\nabla^{\sigma}Ug_{\mu\nu} + 4R_{\mu\rho\nu\sigma}\nabla^{\rho}\nabla^{\sigma}U = 0.
\end{align}
\end{widetext}
The scalar field equation combined with the trace of the above equation leads to
\begin{align}\label{sq}
	3\phi = \mu^2\phi - (1 + \phi)U'(\phi) \mathcal{G} - 2R\Box U(\phi) + 4R_{\mu\nu}\nabla^{\mu}\nabla^{\nu}U.
\end{align}
When \( U = 0 \), the equation describes a standard free massive scalar. It is crucial that the scalar equation in the extended theory is algebraic, allowing it to be integrated out to yield pure gravity \cite{Liu}. Considering \( U = \frac{\beta \phi^n}{n} \), the scalar field equation (\ref{sq}) gives \( \phi = 2\alpha (R + \beta \mathcal{G}) \) for \( n = 1 \), \( \phi = \frac{2 \alpha R}{1 - 2 \alpha \beta \mathcal{G}} \) for \( n = 2 \), and \( \phi = \frac{3 - \sqrt{9 - 192 \alpha^2 \beta R \mathcal{G}}}{8\alpha \beta \mathcal{G}} \) for \( n = 3 \). Hence, the Lagrangians (\ref{lag}) can be expressed as
\begin{align}\label{lag1}
	\mathcal{L} = \sqrt{-g} \left(R + \alpha(R + \beta \mathcal{G})^2\right), \hspace{1cm} n=1,
\end{align}
\begin{align}\label{lag2}
	\mathcal{L} = \sqrt{-g} \left(R + \frac{\alpha R^2}{1 - 2 \alpha \beta \mathcal{G}} \right), \hspace{1cm} n=2,
\end{align}
\begin{align}\label{lag3}
	\mathcal{L} = \sqrt{-g} \bigg(R &+ \frac{3R}{32 \alpha \beta \mathcal{G}} - \left(\frac{3 - \sqrt{9 - 192 \alpha^2 \beta \mathcal{G} R}}{1024 \alpha^3 \beta^2 \mathcal{G}^2}\right) \nonumber \\ & \left(3 - 48\alpha^2 \beta R \mathcal{G}\right)\bigg),\quad n=3.
\end{align}
Since we are considering a corrections to the Starobinsky Lagrangian \( \sqrt{-g} (R + \alpha R^2) \),  at the small-\(\beta\) limit, the GB-corrections to the above Lagrangians lead to \( 2 \alpha \beta R \mathcal{G} \), \( 2 \alpha^2 \beta R^2 \mathcal{G} \), and \( \frac{8\alpha^3 \beta R^3 \mathcal{G}}{3} \), respectively. We know that \( \alpha R^2 \) is a correction to the Einstein Lagrangian itself. Therefore, choosing \( n=1 \) we can consider the first-order correction as  
\begin{equation} \label{lagx}
	\mathcal{L} = \sqrt{-g} (R + \alpha R^2 + 2 \alpha \, \beta \, R \, \mathcal{G}).
\end{equation}
The above equation shows that the RGB model is the first-order extended GB-correction to the Starobinsky Lagrangian, where we have $\text{RGB}=R \, \mathcal{G}$, and the coupling constant $2 \alpha \beta$ can be considered as a new $\beta$ in Eq. (\ref{LRGB}).

\clearpage
\widetext
\end{document}